\documentclass[12pt]{article}
\usepackage[utf8]{inputenc}
\usepackage{graphicx,psfrag,epsf}
\usepackage{booktabs}
\usepackage{textgreek}
\usepackage{threeparttable}
\usepackage{float}
\usepackage{amsmath,amsfonts,amsthm,bm} 
\usepackage{url}
\usepackage{transparent}
\usepackage{adjustbox}
\usepackage{flafter}
\usepackage{lscape}
\usepackage{caption}
\usepackage{subcaption}
\usepackage{graphicx}
\usepackage{geometry}
\usepackage{footnote} 
\makesavenoteenv{tabular} 
\usepackage{verbatim}
\usepackage{natbib}
\usepackage{comment}
\usepackage{rotating}
\usepackage{hyperref}
\usepackage{mdframed}
\usepackage{lipsum}
\usepackage{array}
\newcolumntype{H}{>{\setbox0=\hbox\bgroup}c<{\egroup}@{}}
\usepackage{xcolor}
\usepackage{amsmath}
\usepackage{multirow}
\usepackage{bbm}
\usepackage{algorithmicx}
\usepackage{algorithm,algpseudocode}
\usepackage{longtable}

\newcommand{\tibi}[1]{\textcolor{blue}{$<<$TS: #1$>>$}}

\newcommand{\blind}{0}

\begin{document}

\def\spacingset#1{\renewcommand{\baselinestretch}%
{#1}\small\normalsize} \spacingset{1}

\def\spacingset#1{\renewcommand{\baselinestretch}%
{#1}\small\normalsize} \spacingset{1}

\newcommand{\papertitle}{Macroprudential Policy and Downside Risk: Regime-Dependent Effects of Capital Regulation}

\if0\blind
{
  \title{\bf \papertitle\thanks{
    The authors thank Tam\'as Bork\'o, Zita Fellner, and Vivien K\'ad\'ar-Vir\'agh for providing assistance with the data and comments on early drafts. We are thankful to the participants of the 2026 RCEA conference for their feedback and questions. Early versions of the paper were presented under the title ``Estimating Short and Long Term Effects of Capital Requirements - Evidence from Hungary'' The usual disclaimer applies.}}
  \author{
    Vivien Czofa\\
    Central Bank of Hungary, Hungary\\
    \\
    Tibor Szendrei\\
    National Institute of Economic and Social Research, UK.\\
    \\
    Katalin Varga\footnote{Corresponding author: vargaka@mnb.hu.}\\
    Central Bank of Hungary, Hungary.}
  \maketitle
} \fi

\if1\blind
{
  \bigskip
  \bigskip
  \bigskip
  \begin{center}
    {\LARGE\bf \papertitle}
\end{center}
  \medskip
} \fi

\begin{center}
\end{center}

\begin{abstract}
\noindent This paper employs a Threshold Bayesian Vector Autoregression (TBVAR) to estimate the regime-dependent macroeconomic effects of capital regulation in Hungary. Using the Factor-based Index of Systemic Stress (FISS) as the threshold variable, the model identifies normal and stress regimes consistent with the occasionally binding constraints literature. The TBVAR offers a practical multivariate alternative to Growth-at-Risk for data-constrained economies. Generalised impulse responses reveal a pronounced asymmetry: releasing regulatory capital during stress raises GDP growth at the peak, with effects persisting for roughly twenty months, while the cost of accumulating capital in the normal regime is economically negligible. These findings are robust to alternative Cholesky orderings, sample periods, and credit variable definitions, providing direct empirical support for the countercyclical operation of the capital buffer.
\end{abstract}


\noindent%
\noindent
\vfill

\spacingset{1.45} 


\section{Introduction}
The global financial crisis demonstrated that macro-financial linkages are fundamentally nonlinear: financial distress does not merely reduce expected growth but disproportionately increases the probability of severe contractions \citep{brunnermeier_macroeconomic_2014}. \citet{adrian_vulnerable_2019} formalised this insight through the Growth-at-Risk (GaR) framework, which uses quantile regression to show that deteriorating financial conditions compress the lower quantiles of GDP growth while leaving the upper quantiles largely stable. The framework has since become a mainstay of macroprudential analysis, underpinning the stance assessment of several European authorities \citep{skrinjaric_impact_2025,european_systemic_risk_board_improvements_2024} and informing policy work at Central Banks \citep{chavleishvili_risk_2021}.

Despite its influence, GaR faces two practical limitations. First, it is predominantly univariate, whereas macroeconomic systems are characterised by pervasive endogeneity best addressed through multivariate modelling \citep{sims_macroeconomics_1980}. Quantile VAR extensions exist \citep{chavleishvili_forecasting_2024}, but their estimation requires either a pre-specified recursive ordering that becomes difficult to validate as the number of variables grows, or computationally intensive optimal transport methods \citep{carlier_vector_2016}. Second, quantile estimation is data-intensive, particularly in the tails where macroprudential interest is concentrated. The combination of limited observations and the univariate nature of GaR renders the approach impractical for many small-country applications.

This paper proposes a Threshold Bayesian VAR (TBVAR) as a practical alternative for studying the regime-dependent macroeconomic effects of macroprudential capital regulation. The approach has both empirical and theoretical motivation. \citet{caldara_understanding_2020} show that a Markov-switching model can replicate the key GaR finding of a negative mean–volatility correlation in future growth, while \citet{carriero_capturing_2024} demonstrate that Bayesian VARs with time-varying volatility capture macroeconomic tail risks. Theoretically, the threshold specification maps onto models with occasionally binding borrowing constraints, where a financial accelerator amplifies shocks only when constraints bind \citep{brunnermeier_macroeconomic_2014,he_model_2012}. \citet{aruoba_piecewise-linear_2021,aruoba_svars_2022} establish that the decision rules from such models are approximately piecewise-linear, justifying a VAR with regime-dependent coefficients. Following \citet{alessandri_financial_2017}, we use a factor model capturing the degree of stress in the financial system as the threshold variable that proxies the unobservable tightness of borrowing constraints.

The policy question at the centre of this paper is whether the countercyclical capital buffer (CCyB) operates as intended: accumulated at low cost during expansions and released during stress to sustain credit supply and mitigate output losses \citep{bank_of_england_financial_policy_committee_financial_2023,aikman_would_2019}. Micro-level evidence from Spain’s dynamic provisioning framework shows that countercyclical buffers can sustain credit supply during downturns \citep{jimenez_macroprudential_2017}, but the aggregate macroeconomic consequences of buffer release (particularly for small open economies) remains an open question. We address this by estimating regime-dependent impulse responses of GDP and credit to regulatory capital shocks.

We apply the TBVAR to Hungary, which is an instructive case study: its limited data history tests the constraints under which GaR methods operate; it is a small open economy with functioning financial markets; and it has developed a Factor-based Index of Systemic Stress (FISS), a composite financial stress indicator methodologically comparable to the NFCI used in the US GaR literature. (Using the FISS as the threshold variable also has direct policy justification, since CCyB release decisions are typically predicated on financial stress indices \citep{detken_operationalising_2014}.)
Using the FISS as the threshold variable also has a direct policy justification.  In practice, large or discretionary releases of the countercyclical capital buffer  (CCyB) are often associated with episodes of elevated financial stress and are  therefore closely monitored using financial stress indicators \citep{detken_operationalising_2014}.  At the same
time, more gradual reductions in the buffer may reflect developments in the credit cycle rather than acute financial distress. Future research could therefore consider jointly modelling financial stress and credit-cycle indicators (for example CRI measures) or allowing for multiple threshold variables within the model.

We have two key findings. First, generalised impulse responses reveal a pronounced asymmetry: releasing regulatory capital during stress raises GDP growth by on standaard deviation at the peak, with the effect persisting for roughly twenty months, while the cost of building up capital in the normal regime is economically negligible. This asymmetry is robust to alternative Cholesky orderings, sample periods, and credit variable definitions. Second, the credit response is more sensitive to specifications, suggesting that transmission operates through multiple channels beyond the volume of new lending. The paper thus contributes to the nonlinear macro-financial VAR literature \citep{alessandri_financial_2017}, complements micro-level CCyB evidence \citep{jimenez_macroprudential_2017,dursun-de_neef_countercyclical_2023} with aggregate evidence, and demonstrates that threshold methods offer a practical multivariate alternative to GaR for data-constrained economies.

The remainder of the paper is organised as follows. Section 2 motivates the TBVAR as an approximation to the GaR framework. Section 3 describes the data. Section 4 sets out the methodology. Section 5 presents the results. Section 6 concludes.




\section{From GaR to TVAR: Motivation}
\citet{adrian_vulnerable_2019} has pioneered the use of Growt-at-Risk (GaR) to measure downside risk of GDP growth. The procedure has become a mainstay for macroprudential policy since the objectives of the two align: the idea of GaR is inference on tails, which is precisely where macroprudential policies exert their largest influence \citep{galan_benefits_2024}. Because of this GaR has not only become a popular tool for central bankers (as seen in papers such as \citet{chavleishvili_risk_2021} or \citet{figueres_vulnerable_2020}) but has been adopted as part of the Macroprudential stance of many Central Banks \citep{skrinjaric_impact_2025,european_systemic_risk_board_improvements_2024}.

A key limitation of GaR is that it is currently predominantly a univariate framework, whereas macroeconometrics is rife with endogeneity, which is best tackled with multivariate modelling \citep{sims_macroeconomics_1980}. To fill this gap QVAR methodologies (like \citet{chavleishvili_forecasting_2024}) have been recently proposed but estimation of these methods is not straightforward. The key challenge is how one defines a multivariate quantile. If one opts to use the Rosenblatt transformation for the multivariate quantile definition, then one has to pre-specify an ordering and as the number of variables increases the ordering becomes difficult to ascertain. This sequential ordering is instrumental for estimation and there is no established diagnostic for validating the recursive ordering in the QVAR context. Other multivariate quantile estimation methods that do not require such ordering are based on optimal transport \citep{carlier_vector_2016} methods, which are computationally intensive.

Another key limitation of GaR methods is the data requirements. While GaR is undeniably a powerful tool, its reliance on quantile estimation methods mean that the data requirements can be large especially if one is interested in inference in the tails. This has prompted the development of methods that can recover quantile profiles even in the presence of limited observations \citep{fernandes_smoothing_2021,szendrei_revisiting_2023}. The data limitation coupled with the predominantly univariate use of GaR makes the approach impractical for small-country macro applications.

Given these challenges, how could countries with limited data use multivariate methods that approximate the logic of GaR: i.e. inference about downside. The literature has found evidence that using multivariate or switching methods can recover insight about downside risk. Using threshold methods to approximate GaR has been done in \citet{caldara_understanding_2020} who show that a Markov-switching model with endogenous transition probabilities can replicate the key findings of GaR. Specifically the authors show that a switching AR model yields negative correlation between conditional mean and conditional volatility of future growth. \citet{carriero_capturing_2024} also highlight that Bayesian VARs with stochastic volatility can capture macroeconomic tail risks. As such multivariate methods such as a VAR with threshold effects can approximate the fitted values of QR. Furthermore, VARs can be used for impulse response function analysis, which allows for the testing of key policy questions. As such, we propose using a Threshold VAR approach to approximate the GaR framework. We will then test the release of regulatory capital in the stress regime to test what the release of the Countercyclical Capital Buffer (CCyB) would be on the aggregate economy.

Hungary is used as a case study for the Threshold VAR application. It is an ideal candidate for three reasons: (1) its limited data history directly tests the constraints under which GaR methods operate; (2) it is a small open economy; and (3) it possesses functioning financial markets, albeit ones that remain relatively underdeveloped due to the absence of deep derivative markets \citep{szendrei_fiss_2020}. The presence of functioning financial markets along with the openness of the economy make Hungary an ideal candidate, as these properties make the nonlinear macro-financial linkage emerge, which is precisely what the GaR is designed to capture. Furthermore, Hungary has a favourable data environment. In particular, Hungary has a well-established financial stress index (the Factor Based Index of Systemic Stress (FISS) of \citet{szendrei_fiss_2020}) that is methodologically similar to the National Financial Conditions Index (NFCI) used in the US GaR literature. Indices of this type aggregate information across a broad set of financial variables, capturing the prevailing level of financial stress (or financial conditions) rather than the changes in financial stress. As such, these measures are more likely to be slow to return to ``normalcy'' after a spike in financial stress, which means it also tracks the aftermath of financial shocks. This makes them suitable for conditioning variables in the GaR framework \citep{adrian_vulnerable_2019} and threshold variable in the Threshold VAR setting \citep{alessandri_financial_2017,alessandri_financial_2019}. An important distinction between the NFCI and the FISS is that while the NFCI is a financial conditions index the FISS is a financial stress index. This means that while the NFCI has information pertaining to the full conditional distribution of GDP, the FISS only carries information that is likely to be concentrated in the left tail of the target distribution.

We will use the FISS as the variable that determines the regimes. This is motivated by two reasons. The first reason is that a threshold VAR specification can be motivated by the occasionally binding constraints literature. Specifically, the financial stress indicator can be used as a threshold variable that proxies the unobservable tightness of borrowing constraints \citep{alessandri_financial_2017}. In models with occasionally binding constraints, borrowing constraints may or may not bind at a given point in time. When the constraint is slack, the economy behaves as a standard model would predict, but when it binds, a financial accelerator mechanism amplifies the propagation of shocks. This binding constraint in turn can have dramatic consequences for output \citep{brunnermeier_macroeconomic_2014,he_model_2012}. \citet{aruoba_piecewise-linear_2021} and \citet{aruoba_svars_2022} show that the decision rules arising from such models are approximately piecewise-linear, so one can recast the model as a VAR with regime-dependent coefficients, that switch when the constraint becomes binding.  

The second reason for using the FISS as a threshold variable is that the CCyB release is often based on stress indices \citep{detken_operationalising_2014}. Using the FISS as the threshold variable ensures that the stress regimes identified by the model correspond to the episodes where the Macroprudential authority would release the CCyB. This allows us to study the impacts of CCyB release on GDP growth. As such, using the FISS as the threshold variable has both theoretical and policy justification.

\begin{figure}[!t]
    \centering
    \includegraphics[width=\linewidth]{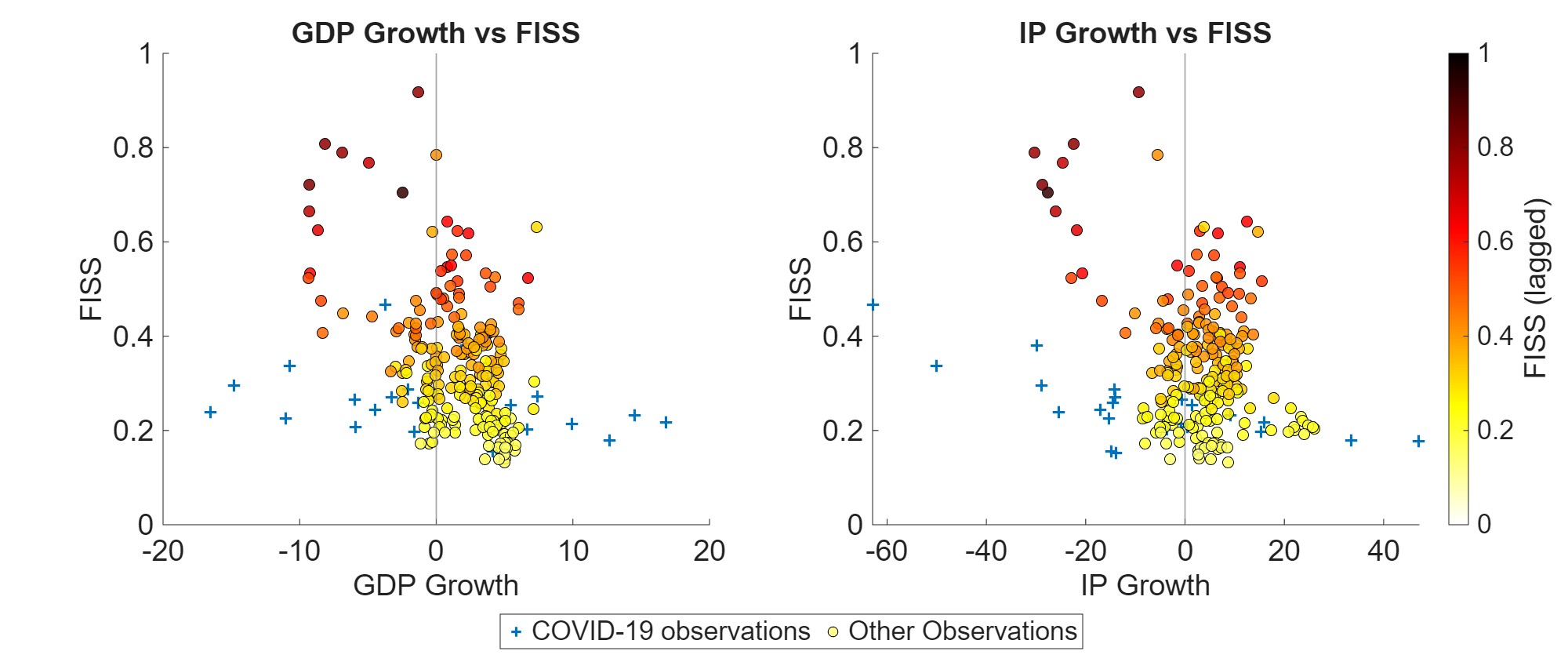}
    \caption{Scatterplot of FISS and growth}
    \label{fig:Scatter}
\end{figure}

While the choice of the FISS as threshold variable is well motivated, it is worth examining whether the empirical evidence corroborates the justifications raised above. Figure \ref{fig:Scatter} presents a scatter plot of constructed monthly GDP\footnote{Details on how the Monthly GDP was constructed is found in the appendix.} year-on-year growth against the FISS, with observations colour-coded by the lagged level of the stress index.\footnote{The lag is set to 1, which is the optimal delay parameter estimated in the Bayesian Threshold VAR.} The figure reveals that as one moves towards the upper-left quadrant (i.e., towards higher financial stress and more negative output growth) the colour gradient shifts systematically from yellow to red. This indicates that elevated lagged stress is concentrated among contractionary episodes. The asymmetry is consistent with the occasionally binding constraints view, in which the financial accelerator is active only when stress is elevated, and provides empirical support for the use of the FISS as a regime-determining variable. In this figure the COVID-19 observations are displayed separately. This is because the COVID-19 era observations are characterised by large swings in output growth that were lockdown measure driven rather than attributable to financial distress. 

To assess whether this pattern is an artefact of the monthly GDP estimation procedure, Figure \ref{fig:Scatter} also plots the Industrial Production year-on-year growth against the FISS. The same relationship is present which alleviates the concern that the relationship is an artefact of the monthly GDP estimation procedure. Since IP is a directly observed series, its co-movement with the lagged FISS reflects a structural relationship between financial conditions and real activity rather than a feature of the interpolation methodology used to construct the monthly GDP estimate.

\begin{figure}[!t]
    \centering
    \includegraphics[width=\linewidth]{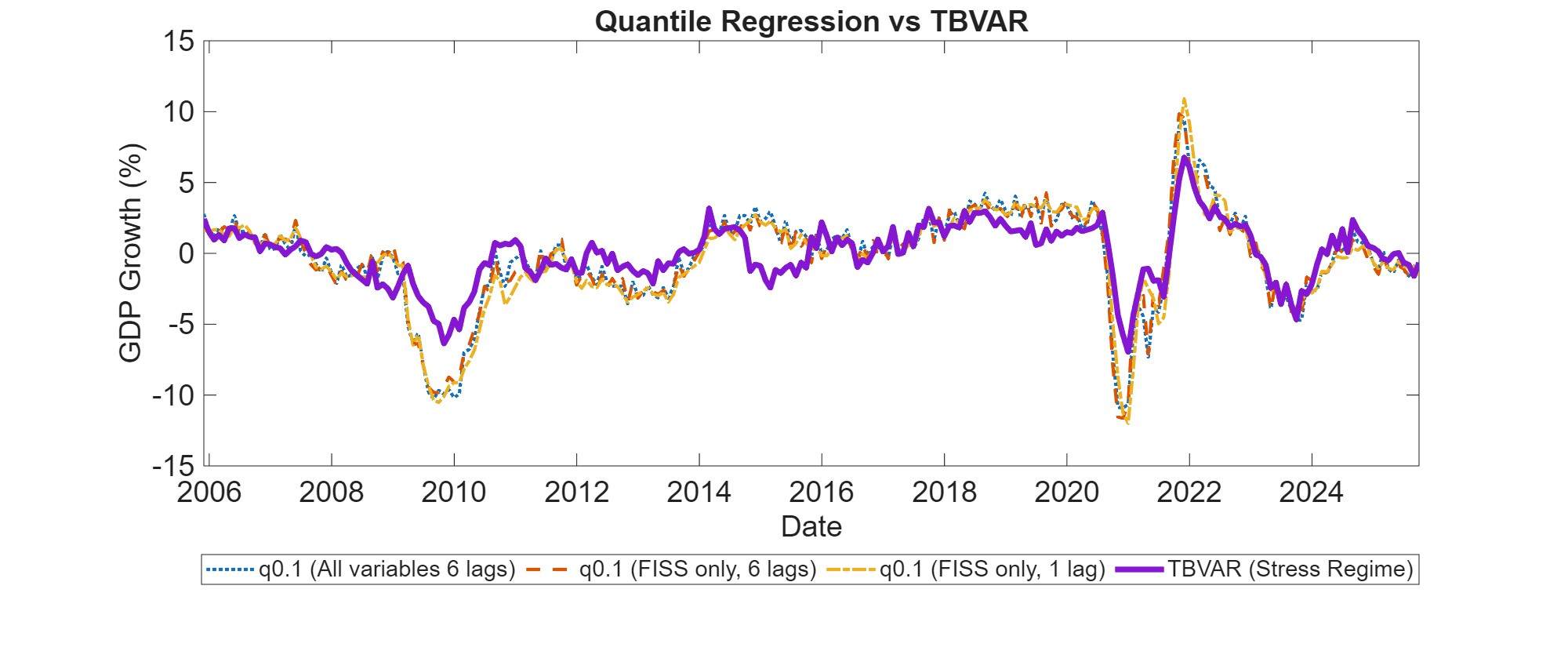}
    \caption{TBVAR Stress Regime and QR ($10^{th}$ quantile) fits of Monthly GDP}
    \label{fig:StressFit}
\end{figure}

We also present the fitted values of the Stress Regime VAR and various Quantile Regression (QR) fits in figure (\ref{fig:StressFit}). To construct these quantile regression fits, the procedure of \citet{szendrei_revisiting_2023} was used, which imposes non-crossing constraints and an adaptive LASSO shrinkage on the levels of the variables. 3 QR specification were ran, the Canonical GaR which has lagged GDP and FISS only with 1 lag; an extended GaR which has 6 lags of FISS and GDP; and a Full GaR which has the same variables as our proposed TBVAR with the same number of lags.\footnote{Because of the number of parameters in the last GaR, imposing shrinkage on the coefficients is necessary.} For all QR models, every $10^{th}$ quantile was estimated (9 quantiles in total), but we only look at the similarity between downside fits and the Stress Regime of the TBVAR. This is very similar to the exercise \citet{caldara_understanding_2020} do to showcase how one can use the stress regimes of a Markov-Switching VAR to recover the the left tail of a GaR. We note that while \citet{caldara_understanding_2020} focus on the stress regime of a MS-VAR and the $25^{th}$ quantile of a GaR, we will be looking at the stress regime of a TBVAR and compare it to the $10^{th}$ quantile of a GaR. We opt to focus on the lower $10^{th}$ quantile, because Central Banks are more likely to be interested in the extreme left tail ($5^{th}$ and $10^{th}$) of GDP growth rather than an intermediate one \citep{aikman2019credit,franta2020effects}.

The figure reveals a broad agreement in the dynamics of the four series: between major downturns, the quantile regression fitted values and the TBVAR stress-regime path align closely, tracking the same periods of moderate downside risk with comparable amplitude. This co-movement between QR measures and the TBVAR Stress regime is indicative of the TBVAR to approximate downside risk. However, this is merely an approximation as we can see that during the downturns (such as the 2008-2009 global financial crisis and the COVID-19 contraction of 2020-2021) the QR specifications produce deeper downside risk estimates than the TBVAR stress regime. This reflects a fundamental difference in the two approaches: QR directly targets a specific quantile of the conditional distribution and can therefore capture extreme tail realisations more flexibly, while the TBVAR model conditions on a discrete regime indicator and estimates a single set of parameters governing all observations within the stress state. Consequently, while the TBVAR provides an informative approximation of downside risk dynamics, it cannot (at least for Hungarian data) deliver a one-to-one correspondence with direct quantile regression fitted values. 

Taken together, these results point to the GaR and TBVAR being complementary rather than competing methodologies. While a fully specified quantile VAR (such as \citet{chavleishvili_forecasting_2024}) could deliver both non-linear Impulse Response functions to shocks and accurate downside risk estimates, it requires enough observations along the full joint distribution to identify each quantile level. The TBVAR collapses the multivariate quantile estimation problem into two conditionally linear BVARs rather than requiring separate quantile fits at every point of the distribution. This lowers the data burden required for the model to be estimable and it is capable of approximating downside risk, while giving regime dependent impulse responses. For a policy question such as the macroeconomic effect of releasing the CCyB in Hungary, this is a worthwhile trade-off: it sacrifices some precision in the depth of the tail in exchange for a model that remains estimable in a data-constrained setting such as Hungary. On account of this where data availability permits, quantile VAR methods remain the preferred framework. Nevertheless, for economies with limited data history the TBVAR offers a practical alternative.

\section{Data}

In the macro-financial linkages literature it is customary to use financial stress or financial condition indices.\footnote{Bond spreads are also a commonly used variable. However, these measures are not as encompassing as financial stress and financial conditions indicators.} In this paper we will follow this by using a financial stress indicator as a threshold variable. Specifically we will use the Factor-based Index of Systemic Stress (FISS), developed by \citet{szendrei_fiss_2020}. The FISS is a continuous, daily measure of contemporaneous financial stress in the Hungarian financial system, scaled to the unit interval. It is constructed using a dynamic Bayesian factor model that compresses 19 high-frequency financial variables that span the government bond market, foreign exchange market, capital market, and the bank and interbank segment. A key advantage of the FISS is that its non-stationary factor structure captures the level of financial stress rather than merely changes in stress, which makes it particularly well-suited as a threshold variable in a regime-switching framework. Specifically, because the index measures the level of stress it is more likely to identify consecutive periods as regimes rather than single instances when stress spiked. The generality of the underlying methodology has been demonstrated by \citet{varga_non-stationary_2025}, who apply the same framework to construct a financial stress index for the United Kingdom, showing that it performs well in a substantially different market environment.

The credit variable used is a flow measure: monthly new credit extended by Hungarian banks, deflated by core CPI. The underlying series aggregates new lending denominated in Hungarian forint, Swiss francs, and euros, and covers both mortgage-secured household credit and lending to non-financial corporations (including consumer loans in the baseline specification). A flow measure is preferred over the year-on-year change in credit stocks for two reasons. First, monthly new lending captures the contemporaneous supply of credit to the economy, which is the margin most directly affected by changes in bank capital requirements \citep{jimenez_macroprudential_2017,dursun-de_neef_countercyclical_2023}. Second, in our specification the differenced credit stock variable produced substantially inferior model fit relative to the flow measure, consistent with the observation that stock-based measures can inherit low-frequency trends that obscure the higher-frequency transmission of policy shocks.

We also estimate the model using an alternative credit flow that excludes unsecured consumer lending (personal loans, credit cards, overdraft facilities, and goods-purchase credit). This narrower measure retains only mortgage-secured household credit and lending to non-financial corporations. By doing this, we isolate the components of new lending most directly linked to collateral cycles and productive investment. The motivation for this decomposition is twofold. First, housing markets and financial stability are deeply interconnected \citep{szendrei_roof_2026}. Specifically, mortgage-secured lending constitutes the bulk of household (and many banks’) credit exposure, and fluctuations in collateral values create a feedback loop between house prices, borrowing capacity, and the real economy \citep{iacoviello_house_2005}. Second, when banks face a tightening of regulatory capital requirements, theory and evidence suggest that they are more likely to curtail unsecured consumer lending (which carries higher risk weights under the Basel framework and is easier to adjust at the margin) while preserving mortgage-secured and corporate lending, which is collateralised and central to banks’ long-term balance-sheet strategies \citep{dursun-de_neef_countercyclical_2023,auer_countercyclical_2022}. By extension if the CCyB is released during stress, the resulting increase in lending capacity should flow primarily into the secured and corporate channel through which the buffer is designed to operate. Comparing the impulse responses of this alternative credit measure against those from the baseline model allows us to assess whether the transmission of regulatory capital shocks operates primarily through the secured and corporate lending channel, as the financial stability rationale for the CCyB would suggest.

We note that the FISS is preferred as our financial stress indicator in part because it does not include credit variables in its construction. Since credit enters the VAR as a separate endogenous variable, using a financial stress index that also contained credit components would effectively double-count credit information in the system. The exclusion of credit from the FISS ensures that the two variables capture distinct dimensions of financial conditions, allowing us to cleanly identify the impulse responses of credit to shocks within each regime.

As a control variable the broad Real Effective Exchange Rate (REER) calculated by the BIS is also included. In Hungary, credit denominated in foreign currency constituted a significant share of banks' loan portfolios until the end of 2014, when a government-mandated conversion programme eliminated most foreign currency household loans. Because of this, exchange rate movements had a direct impact on the domestic value of outstanding credit (and borrowers debt-servicing capacity) during part of the estimation period. As such, including the REER is important to account for this channel of the model.

To capture the solvency capital position of the Hungarian banking system, we include two complementary variables: regulatory capital requirement and 
free or voluntary capital buffers, both expressed as proportions of total-risk-weighted assets (RWA).\footnote{We will refer to total risk-weighted-asset as RWA. In the European context this is equivalent to TREA}
Regulatory capital is defined as the sum of Tier 1 capital, all first- and second-pillar as well as combined buffer requirements imposed on Hungarian banks. 
Pillar 1 capital requirements are the mandatory minimum capital standards under the Basel Accord, requiring banks to hold capital equal to at least $8\%$ of their RWA for credit, market, and operational risks. Pillar 2 capital requirements are bank-specific, legally binding capital add-ons imposed by supervisors through the Supervisory Review and Evaluation Process (SREP) to cover risks not fully captured by minimum (Pillar 1) requirements, such as business model, governance, and interest rate risks. Combined buffer requirements are mandatory capital layers above minimum regulatory requirements, designed to ensure banks hold sufficient solvency capital to absorb losses. These additional requirements include the Capital Conservation Buffer (CCoB), Countercyclical Buffer (CCyB), and Systemic Risk Buffer (SyRB) and buffers for systemic importance (G-SII/O-SII).

In Hungary, CET1 capital accounts for the overwhelming majority of banks’ capital structures: a reflection of both high profitability in the sector and the relative complexity and cost of issuing subordinated instruments.\footnote{We note that until 2014, the regulatory capital requirement for the Hungarian banking system was effectively constant at 8\% of RWA (with minor deviations) and was reported at monthly frequency in the Common Reporting (COREP) framework. From 2014 onward, reporting shifted to quarterly frequency and the regulatory capital series became non-constant, reflecting the introduction of second-pillar SREP add-ons and other macroprudential buffers. To obtain a monthly series for the full sample, we linearly interpolate the quarterly regulatory capital observations over the post-2014 period. Both capital variables are transformed to year-on-year growth rates to ensure stationarity and consistency with the other variables in the system.} As such, the overall capital adequacy constraint is the binding margin rather than the composition of capital tiers.

Free capital is defined as solvency capital minus regulatory capital (pillar 1 and pillar 2 combined) buffer requirements expressed relative to RWA. It represents the buffer held voluntarily by the banking system above the regulatory minimum. While regulatory capital responds primarily to policy decisions (such as changes to the different macroprudential buffers or revisions of SREP requirements) free capital buffers reflects the aggregate of bank-specific management decisions regarding leverage, risk appetite, and balance-sheet strategy. We include both free and regulatory capital together since these two variables capture total solvency capital in the banking system. 




These variables capture different aspects of risk and as such we include them as individual variables rather than an aggregate of the measures. If we were to only include the bank solvency capital variable, a shock to this variable would conflate prudential policy changes with endogenous adjustments in banks' own capital management. We note that omitting free capital entirely is not advised as that would discard important information: bank equity tends to be more volatile than regulatory requirements alone, and movements in free capital in the banking system capture shifts in the banking systems leverage that may have macroeconomic consequences. By including both variables in our model, we can partial out aggregate capital dynamics from regulatory capital, allowing us to interpret shocks to the latter as more closely resembling exogenous changes in prudential policy.


To capture real economic activity we construct a monthly GDP variable. While Industrial production is a frequently used monthly proxy for GDP, the strength of this proxy is not uniformly strong for every country. For instance, for the case of Hungarian data, IP and GDP is less correlated than it usually is in the literature. One can follow the approach of \citet{adrian_vulnerable_2019}, \citet{figueres_vulnerable_2020}, or \citet{szendrei_revisiting_2023} and do the analysis on quarterly values. However, financial markets move at a faster pace then the real economy, and restricting the analysis to quarterly frequency would potentially throw away too much information especially regarding the regimes. As such we instead opt to construct a monthly GDP measure, which has been an active area of research \citep{koop_nowcasting_2021,koop_reconciled_2023,huber_nowcasting_2023,schorfheide_real-time_2021}. Specifically, we use the methodology of \citet{koop_reconciled_2023} to construct Hungarian Monthly GDP measures. Details about the methodology of the monthly GDP are found in the appendix.


\section{Methodology}

\subsection{Threshold VAR}
The Threshold VAR model on the vector endogenous variables $Y_t$ is defined as follows:

\begin{equation}
    Y_t=\Big[c_N+\sum^M_{j=1}B_{N,j}Y_{t-j}+\Sigma^{1/2}_{N,t}\Big]I(z_{t-d}\leq z^*)+\Big[c_S+\sum^M_{j=1}B_{S,j}Y_{t-j}+\Sigma^{1/2}_{S,t}\Big]I(z_{t-d}> z^*)
\end{equation}

\noindent where the model allows for two regimes (denoted by the subscript N and S for normal and stress regimes respectively) based on the level of $z_{t-d}$ relative to an unknown threshold $z^*$. The delay parameter and the optimal threshold are all estimated in the model. Note that $Y_t$ is a vector of variables containing GDP, REER, Credit, Free- and Regulatory Capital, and the FISS. Notably, the FISS enters $Y_t$ both as an endogenous variable and as the threshold variable $z_t$. In this way, the size and propagation mechanism of shocks are allowed to change depending on the level of financial stress. Note that the regime-specific parameters $\{c_i,B_{i,j},\Sigma_i\}$ (where $i\in\{N,S\}$) in the threshold VAR can be interpreted as reduced-form counterparts of the distinct sets of the first-order conditions that arise in DSGE models with occasionally binding constraints. As such, when the index exceeds the critical level $z^*$, the model switches into a ``crisis'' regime where the constraints are binding, changing both the size and propagation of shocks. Note that this formulation is more flexible that and particular structural model, i.e. all VAR parameters are free to differ between the regimes, nevertheless the model will capture the essential nonlinearity that theory predicts should matter. Both regimes are estimated with 6 lags, capturing half a year's worth of history.

We note that the FISS is a factor model and is included as an endogenous variable in the model. As such by its inclusion we augment the information set of the VAR in the spirit of the factor-augmented VAR \citep{bernanke_measuring_2005}. As \citet{alessandri_financial_2017} note, including such indicators turns a small-scale VAR into a model that implicitly conditions on a much larger set of financial variables. This mitigates the well-known concern that small VARs may suffer from omitted variable bias. The predictive content of individual financial variables can shift over time and relying on a broad composite indicator rather than a single series reduces the risk that the results are driven by the idiosyncratic behaviour of a particular variable in a specific stress episode. As such the FISS serves a dual role: it is both an additional endogenous variable that enriches the information set and it is the threshold variable that governs the switch between regimes.

\subsection{Bayesian Estimation}
The posterior distribution of the unknown parameters is simulated using a Gibbs sampler. For this we follow the approach introduced by \citet{chen_bayesian_1995} for threshold autoregressive models which has been extended to the multivariate setting by \citet{alessandri_financial_2017}.

The key insight underlying the estimation strategy is that, conditional on the threshold value $z^*$ and the delay parameter $d$, the regime indicator $I(z_{t-d}>z^*)$ is fully determined. The TB-VAR then reduces to two separate Bayesian VARs, one for each regime. Within each regime $i \in \{N,S\}$, the observations $Y^*_i$ and the corresponding regressor matrices $X^*_i$ are collected, and the conditional posteriors for the VAR coefficients and the error covariance matrix take the standard natural conjugate form \citep{alessandri_financial_2017}:
\begin{equation}
    B_i|\Sigma_i \sim N\!\left(B^*_i,\;\Sigma_i \otimes (X^{*\prime}_i X^*_i)^{-1}\right),~i\in\{N,S\}
\end{equation}
\begin{equation}
    \Sigma_i^{-1}|B_i \sim \mathcal{W}\!\left(Q^*_i,\;T^*_i\right),~i\in\{N,S\}
\end{equation}
where $B^*_i=(X^{*\prime}_i X^*_i)^{-1}(X^{*\prime}_i Y^*_i)$ denotes the regime-specific OLS estimate, $Q^*_i=(Y^*_i - X^*_i \tilde{B}_i)^{\prime}(Y^*_i - X^*_i \tilde{B}_i)$ is the sum of squared residuals with $\tilde{B}_i$ denoting a draw of the VAR coefficients reshaped to be conformable with $X^*_i$, and $T^*_i$ denotes the number of observations assigned to regime $i$. These expressions are standard results in Bayesian VAR estimation under a natural conjugate prior.

The difficulty lies in sampling the threshold $z^*$, since its conditional posterior does not have a known closed-form distribution. To this end we follow \citet{chen_bayesian_1995} and use a random walk Metropolis-Hastings step. Specifically, a candidate value is generated as:
\begin{equation}
    z^*_{new} = z^*_{old} + \psi^{1/2}\varepsilon, \quad \varepsilon \sim N(0,1),
\end{equation}
and is accepted with probability $\alpha = \min\!\left\{1,\; f(Y_t | z^*_{new}, \Theta)\,/\,f(Y_t | z^*_{old}, \Theta)\right\}$, where $f(\cdot)$ denotes the posterior density and $\Theta$ collects all other parameters. A new candidate value of $z^*$ implies a different allocation of observations into the two regimes, so the likelihood must be re-evaluated at each proposal. The scaling factor $\psi$ is calibrated to maintain an acceptance rate between 20\% and 40\%, as is standard practice \citep{chen_bayesian_1995,alessandri_financial_2017}.

For the delay parameter $d$, \citet{chen_bayesian_1995} show that, conditional on all other parameters, its posterior is a discrete multinomial distribution with probabilities:
\begin{equation}
    P_d = \frac{L(Y_t | d, \Theta)}{\sum_{d'=1}^{6} L(Y_t | d', \Theta)},
\end{equation}
where $L(\cdot)$ denotes the likelihood function. This result follows directly from the flat prior imposed on $d$ over the set $\{1, 2, \ldots, 6\}$.

For all TBVAR estimations, we use 15,000 iterations of the Gibbs sampler, discarding the first 10,000 as burn-in and retaining the final 5000 draws for inference. No thinning is applied. The order in which the Gibbs sampler obtains the different parameters per draw is the following:
\begin{enumerate}
    \item \textbf{Regime allocation.} Given the current values of $z^*$ and $d$, determine the regime indicator $I(z_{t-d}\leq z^*)$ and partition the data into the two regimes.
    \item \textbf{VAR coefficients.} For each regime $i\in\{N,S\}$, draw the VAR coefficients $B_i$ from their conditional Normal posterior, given $\Sigma_i$ and the regime-specific data.
    \item \textbf{Error covariance.} For each regime $i\in\{N,S\}$, draw $\Sigma_i^{-1}$ from its conditional Wishart posterior, given $B_i$ and the regime-specific data.
    \item \textbf{Threshold value.} Given the VAR parameters $\{B_i, \Sigma_i\}_{i=N,S}$ and $d$, draw $z^*$ using the Metropolis-Hastings step.
    \item \textbf{Delay parameter.} Given the VAR parameters and $z^*$, draw $d$ from its multinomial posterior.
\end{enumerate}

\subsection{Impulse Response function}


In linear VAR models, impulse response functions are straightforward to compute as they depend only on the model parameters and are invariant to the initial conditions of the system. This is not the case in nonlinear models like the Threshold VAR. As \citet{koop_impulse_1996} demonstrate, impulse responses in nonlinear settings depend on the history of the system at the time of the shock, on the sign and size of the shock itself, and on the composition of the shocks that arrive in subsequent periods. A negative shock hitting the economy during a stress regime may propagate very differently from an identical shock occurring under normal conditions. These features make standard linear impulse response analysis inappropriate for our purposes.

To address this, we adopt the generalised impulse response function (GIRF) framework of \citet{koop_impulse_1996}. The GIRF is defined as the difference between two conditional expectations:

\begin{equation}
    \text{GIRF}(k, \mu_t, \Omega_{t-1}) = E\left[Y_{t+k} \mid \mu_t, \Omega_{t-1}\right] - E\left[Y_{t+k} \mid \Omega_{t-1}\right],
\end{equation}

\noindent where $k$ denotes the forecast horizon, $\mu_t$ is the structural shock of interest, and $\Omega_{t-1}$ is the history of the system up to time $t-1$. The first term represents the expected path of the endogenous variables conditional on both the history and the structural shock, while the second term is the baseline forecast conditional on the history alone. The nonlinear nature of the Threshold VAR model also means that no analytical solution exists for the IRF. As such stochastic simulation of the model is necessary \citep{potter_nonlinear_2000}. The use of GIRFs has become standard practice in the nonlinear macro-financial literature, including applications with threshold VARs \citep{alessandri_financial_2017}, Markov-switching models \citep{hubrich_financial_2015}, and time-varying parameter frameworks \citep{mumtaz_transmission_2015}.

To identify the structural shocks we employ a recursive (Cholesky) decomposition. The ordering of variables is: GDP, REER, Credit, FISS, Free Capital, and Regulatory Capital. By placing Regulatory Capital last (the variable to which we apply the shock) we impose the identifying assumption that innovations to regulatory capital have no contemporaneous effect on any other variable in the system, while regulatory capital itself can respond within the period to developments in all other variables. The real economy variables (GDP and REER) are ordered first, reflecting the assumption that they are the most sluggish and do not adjust within the month to financial or policy innovations. Financial variables (Credit, FISS, Free Capital) are placed in between, as they are assumed to be faster-moving and able to respond contemporaneously to real economy developments but not to regulatory capital changes. This ordering is broadly consistent with the macro-financial linkages literature, where real activity is treated as slow-moving relative to financial conditions \citep{alessandri_financial_2017,chavleishvili_forecasting_2024}. We condition the responses on the observations belonging to each regime separately, which allows us to trace how the propagation of a regulatory capital shock differs between normal and stress periods.

The threshold VAR framework is well-suited to studying the macroeconomic effects of macroprudential capital regulation like the CCyB. Specifically, the CCyB is designed to be released during periods of financial stress, with the aim of easing credit conditions and preventing the financial accelerator from amplifying downturns. If the borrowing constraints bind in the crisis regime but not in normal times, then the effect of easing regulatory capital should be larger when the economy is in distressed state. The threshold VAR allows us to test this directly by estimating regime-dependent impulse responses to a capital regulation shock.

\section{Results}
\subsection{Threshold level}
\begin{figure}
    \centering
    \includegraphics[width=0.75\linewidth]{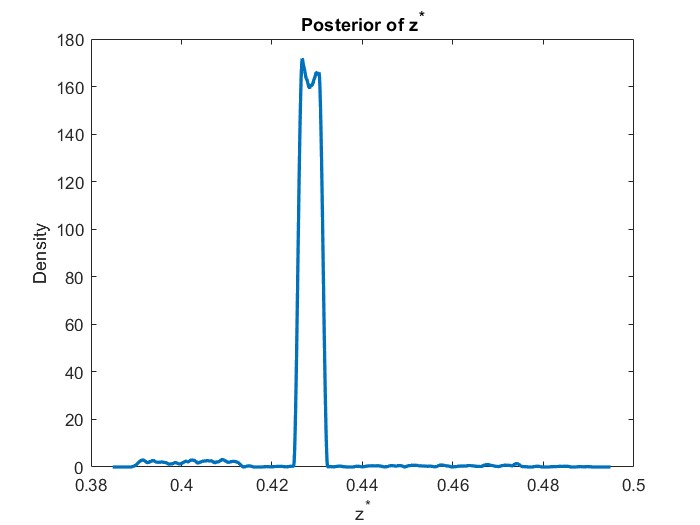}
    \caption{Posterior draws of $z^*$}
    \label{fig:KDFz}
\end{figure}

One advantage of the Threshold BVAR approach is that it identifies the level of the financial stress where the occasionally binding constraints start to bind \citep{alessandri_financial_2017}. At this point the two regimes models parameters start to diverge. Because this level is estimated rather than imposed, the model can be used as a data-driven benchmark for the release phase of the CCyB.\footnote{Note that this is an interpretation of the estimated break rather than a structural identification of the occasionally binding constraint.} As such we start by examining the posterior of $z^*$ which we present found in figure \ref{fig:KDFz} showing the kernel density of draws of $z^*$ obtained jointly with a posterior mode for the delay parameter of 1.

The figure reveals two things. First, the regimes seem to be well identified with a clear spike in the density when $z^*\in[0.42,0.44]$. The bulk of the posterior mass (above 90\%) is in this region. Because the variables are standardised, the corresponding level of FISS where the two regimes' parameters diverge is 0.39, i.e. the regime switches from tranquil to stress regime when the FISS lagged 1 period exceeds a value of 0.39. Second, we can see from the figure that the posterior has a small but noticeable `jitter' outside this interval. We attribute this to the random walk proposal used to sample $z^*$: the chain occaisonally meanders away from the modal region and takes time to return to it. Importantly, these excursions are not merely cosmetic, as each draw in $z^*$ implies a different allocation of observations to the regimes (i.e., the estiamted parameters are influenced as well). In line with \citet{chen_bayesian_1995} we therefore report posterior medians rather than means, which has the added benefit of being robust to the low density draws. We emphasise that while the median is robust to these excursions, outer percentiles used to construct credible intervals are more prone. As such the credible bands in the IRF's below are going to be wider than they would be under a better-mixing sampler. This is pertinent for the stress regime as it is the regime with fewer observations. The implication of this is that any reallocation of observations is a proportionally larger change to the estimation samples.

\subsection{Identified regime}

\begin{figure}[t]
    \centering
    \includegraphics[width=\linewidth]{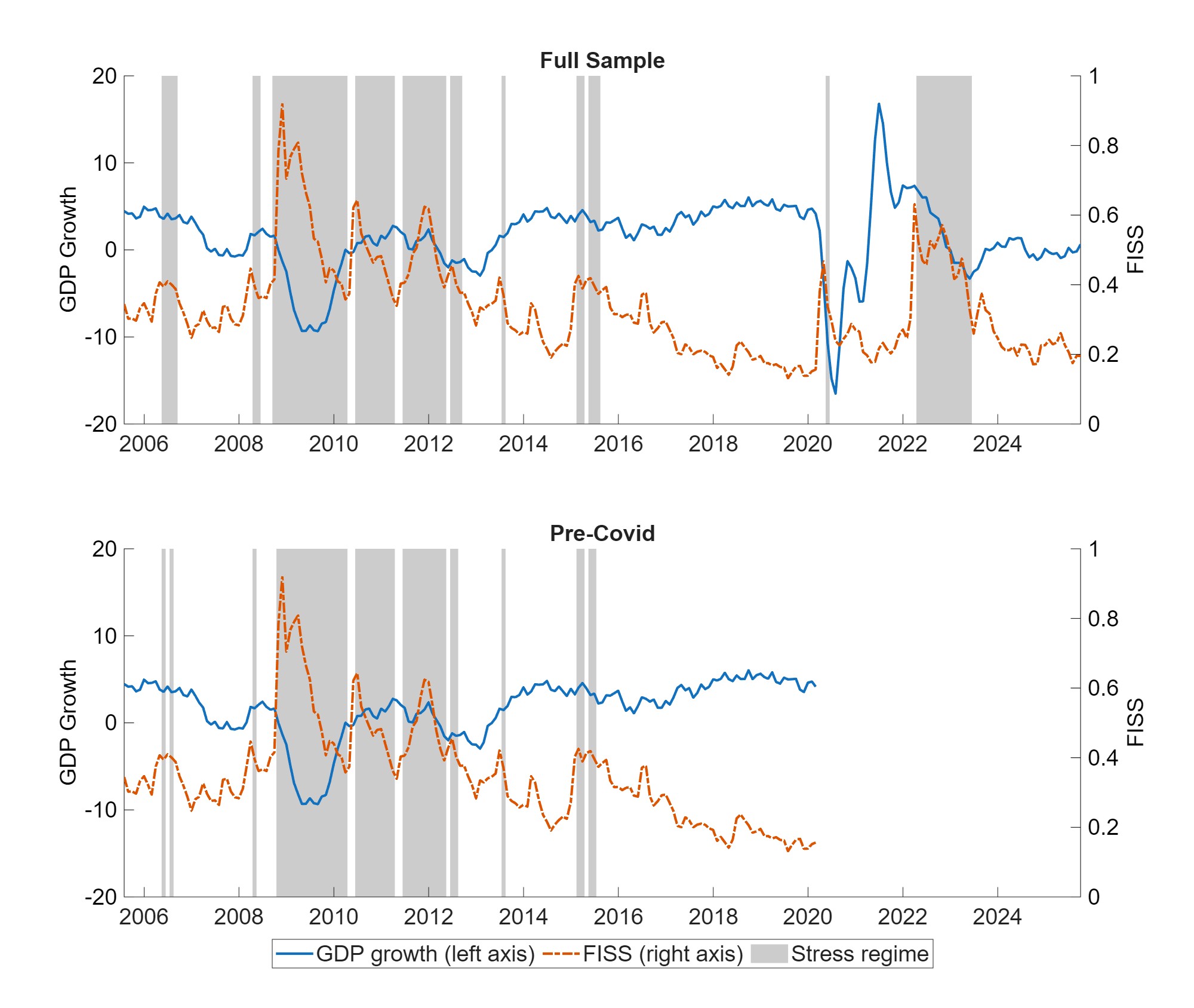}
    \caption{Regimes identified by the Threshold VAR}
    \label{fig:regimes}
\end{figure}

While figure \ref{fig:KDFz} gives an overview of the drawn threshold level, it does not inform us how these regimes look across time. To this end, Figure \ref{fig:regimes} displays the regimes identified by the Threshold VAR for both the full sample and the pre-COVID sample. The grey-shaded areas represent the median posterior estimate of the stress regime indicator, which is equal to one when the FISS exceeds the estimated critical threshold $z^*$. We overlay the regime indicator with the constructed Monthly GDP's year-on-year growth and the FISS.

The full-sample estimation identifies two principal episodes of sustained financial stress. The first corresponds to the global financial crisis (GFC), with the model classifying the period from approximately late 2008 through 2010 as a stress regime, along with several briefer stress episodes in the surrounding years associated with elevated FISS readings during the European sovereign debt crisis and its spillovers to Hungarian financial markets. During this period, GDP growth contracted sharply and the FISS reached its sample maximum, consistent with the severe financial dislocations documented for small open economies during the GFC \citep{brunnermeier_macroeconomic_2014}. The timing and persistence of the identified stress regime are broadly consistent with the pattern reported by \citet{alessandri_financial_2017} for the United States, where the threshold model similarly identifies the GFC as the dominant crisis episode. An important difference, however, is that for Hungary the stress regime extends over a longer window and includes intermittent stress episodes in 2011-2013, reflecting the prolonged impact of the European sovereign debt crisis on Hungarian sovereign spreads and exchange rate volatility.

The second major stress episode identified in the full sample corresponds not to the initial onset of the COVID-19 pandemic in early 2020, but rather to the post-pandemic period from approximately 2022 onward. The model classifies only a brief episode around early 2020 as a stress regime; the sharp contraction in GDP during the first lockdown was driven by public health restrictions rather than by endogenous financial distress, and the FISS, while briefly elevated, returned to moderate levels relatively quickly. By contrast, the sustained stress regime identified from 2022 onward reflects a genuine episode of financial system stress in Hungary. Beginning in late 2021, several adverse shocks created persistent financial stress: surging energy prices following Russia's invasion of Ukraine severely deteriorated Hungary's external balance, with the current account deficit widening to approximately 9 percent of GDP in 2022 \citep{imf2024hungary_articleiv}. The Hungarian currency (forint) depreciated to record lows against major currencies, losing roughly 9 percent against the euro in 2022 alone, driven by the energy price shock, widening fiscal deficits, and uncertainty over the disbursement of EU cohesion and recovery funds \citep{imf2024hungary_articleiv}. Inflation surged to above 25 percent by early 2023, the highest rate in the European Union, prompting the Central Bank of Hungary (MNB) to raise the effective policy rate to approximately 18 percent by October 2022 \citep{imf2024hungary_articleiv}. The MNB's November 2022 Financial Stability Report noted that while the banking sector entered this period with adequate capital and liquidity buffers, the deteriorating economic environment, rising funding costs, and tightening credit conditions posed material risks to financial stability \citep{magyar_nemzeti_bank_financial_2022}. Banks reported plans to tighten lending standards for both household and corporate segments, and the annual growth rate of loan portfolios was expected to decline substantially. The FISS remained at elevated levels throughout this period, reflecting the persistent nature of financial stress during this period. This is precisely what drives the threshold model to classify 2022-2024 as a stress regime.

To verify that the identified stress regimes are not an artefact of the COVID-19 observations, we re-estimate the model on the pre-COVID sample, truncating the data prior to the onset of the pandemic. The bottom panel of Figure \ref{fig:regimes} displays the results. The stress regimes identified in the pre-COVID sample are virtually identical to those in the full sample over the overlapping period. The GFC episode and the intermittent stress periods of 2011-2013 are classified as stress regimes in both estimations, with only marginal differences in their timing or duration: the only difference is the 2006 episode on the full sample is one long stress period, while in the pre-Covid sample it is two smaller stress episodes. This robustness check provides confidence that the regime identification in the full sample is not distorted by the extreme GDP volatility associated with the pandemic-era lockdowns and the subsequent recovery. In particular, the pre-COVID estimation confirms that the threshold value $z^*$ and the delay parameter $d$ are stable across the two sample periods, which implies that the post-2022 stress regime identified in the full sample is classified on the basis of the same FISS dynamics that govern the pre-pandemic regime allocation. The post-pandemic financial stress in Hungary is therefore a genuine stress regime in the sense of the model, driven by the same type of financial market dynamics that characterise the GFC-era stress episodes, rather than being a mechanical consequence of the pandemic-induced output swings.

\subsection{Impulse Response Functions}
\begin{figure}
    \centering
    \includegraphics[width=\linewidth]{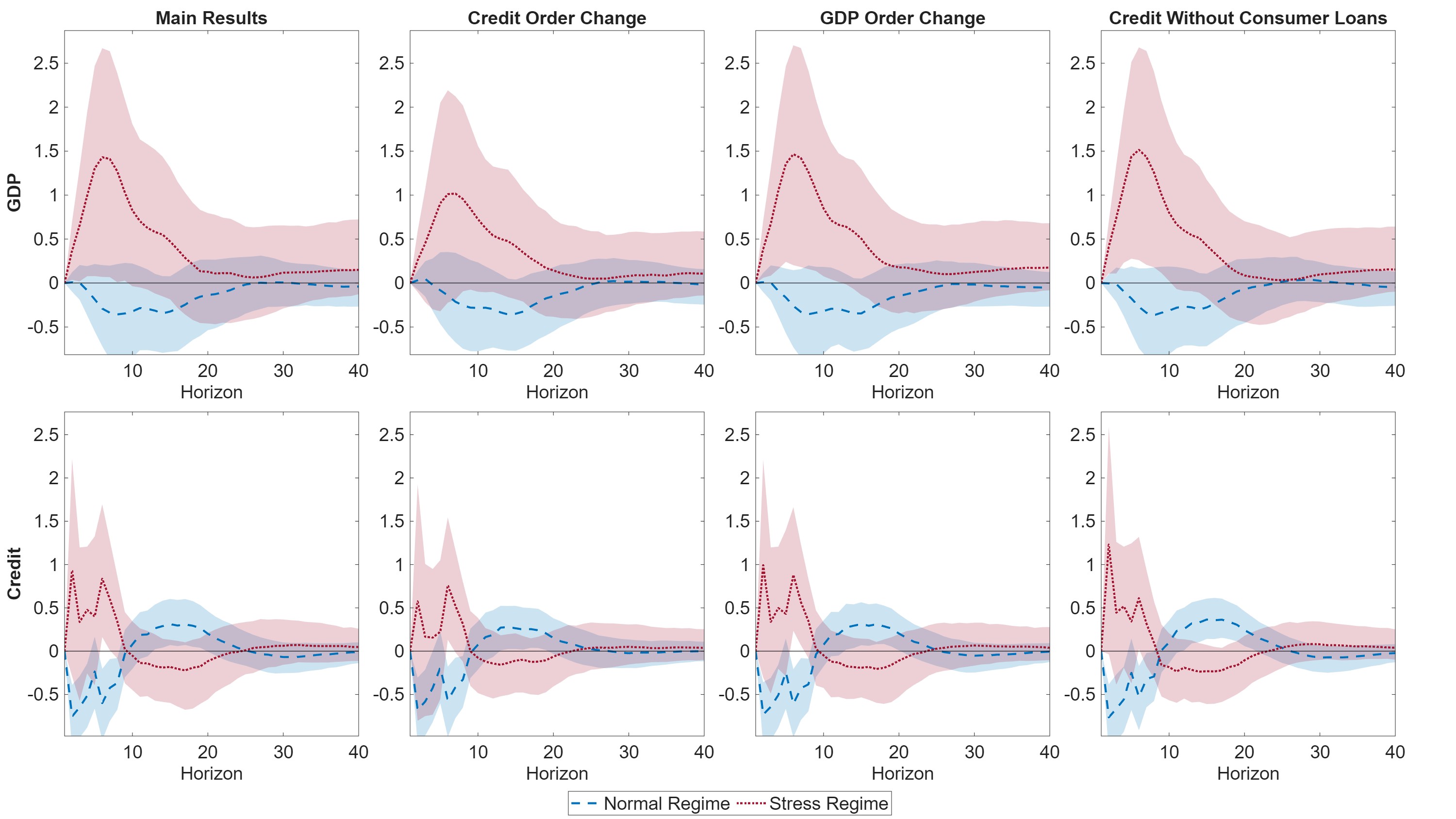}
    \caption{Impulse Responses for the various specifications. Normal regime responses are multiplied by -1 to mimic build-up cost}
    \label{fig:irfsummary}
\end{figure}

All variables have been standardised and as such the IRFs need to be interpreted as standard deviations. Figure \ref{fig:irfsummary} displays the generalised impulse responses of GDP and credit to a one standard deviation change in regulatory capital. The shock is a \emph{decrease} in regulatory capital in the stress regime, which can be interpreted as a release of the CCyB during a period of financial distress; in the normal regime the responses are multiplied by $-1$ so that they can be read as the cost of a one percentage point \emph{increase} in regulatory capital, i.e. the cost of building up the buffer during tranquil times. Median responses and 68\% posterior credible bands are reported for each regime. To assess the sensitivity of the results we present five specifications: the baseline ordering ($Regulatory~Capital\rightarrow Free~Capital\rightarrow Credit\rightarrow FISS\rightarrow REER\rightarrow GDP$); two alternative Cholesky orderings that reposition credit\footnote{$Regulatory~Capital\rightarrow Free~Capital\rightarrow FISS\rightarrow Credit\rightarrow REER\rightarrow GDP$} and GDP\footnote{$Regulatory~Capital\rightarrow Free~Capital\rightarrow Credit\rightarrow GDP\rightarrow REER\rightarrow FISS$} respectively; and the baseline ordering with the narrower credit measure that excludes unsecured consumer lending.\footnote{For the IRF responses we omit generating IRFs on the pre-COVID sample, since the number of observations in the stress regime is significantly lower impeding inference.}

\subsubsection*{GDP response}

The most striking feature of Figure \ref{fig:irfsummary} is the robustness of the GDP response across specifications. In every panel, the median stress-regime response is unambiguously positive: releasing regulatory capital during periods of elevated financial stress raises GDP growth. The response is hump-shaped, rising over the first ten to fifteen months before gradually decaying, a profile that is consistent with the transmission of regulatory capital operating through the credit intermediation channel rather than through an instantaneous expectations channel. 

While Hungary has not yet experienced an actual CCyB release episode the results can be interpreted as capturing the macroeconomic effects of a reduction in regulatory capital requirements more broadly. Under the assumption that a CCyB release operates through the same regulatory capital channel, the  estimated responses provide an indication of the potential macroeconomic impact of such a policy action.

The peak response in the baseline specification reaches approximately 1 standard deviation, and the effect remains positive out to a horizon of roughly 12 months, suggesting that the benefits of CCyB release are not merely temporary demand support but have a more durable effect on the growth path. This persistence is consistent with the view that preventing a credit crunch during stress avoids the hysteresis effects associated with deep downturns \citep{brunnermeier_macroeconomic_2014}.

By contrast, the normal-regime GDP response (shown as the cost of building up regulatory capital) is centred on zero across all specifications. In the baseline, the median hovers around -0.3 standard deviations and the 68\% credible band includes zero throughout the horizon. This near-zero build-up cost is equally robust: it survives changes to the Cholesky ordering, to the sample period, and to the definition of the credit variable.

The combination of a positive release effect with a near-zero build-up cost provides strong empirical support for the CCyB as it is currently envisioned by macroprudential authorities. The countercyclical capital buffer is designed to be accumulated at low cost during expansions and released to sustain lending capacity during downturns \citep{bank_of_england_financial_policy_committee_financial_2023, aikman_would_2019}. Our results suggest that, at least for the Hungarian banking system, the intended asymmetry is present in the data: the cost of higher capital requirements in normal times is economically negligible, while the benefit of lowering them in stress is positive and statistically distinguishable from zero. This finding is consistent with the micro-level evidence of \citet{jimenez_macroprudential_2017} on Spain's dynamic provisioning framework, which shows that countercyclical buffers can sustain credit supply during downturns without imposing material costs during expansions.

It is a standard concern in recursive identification schemes, that the results are exclusively driven by the ordering. Our baseline places regulatory capital last in the ordering, which imposes the restriction that innovations to regulatory capital have no contemporaneous effect on any other variable in the system. This is the most conservative identifying assumption from the perspective of the policy question, because it minimises the within-period impact of the regulatory capital shock on the real economy and financial variables. When we relax this restriction by moving credit later or GDP earlier in the ordering, effectively allowing a more timely response of these variables to the regulatory capital shock, the stress-regime GDP response remains. The Credit Order Change and GDP Order Change specifications yield peak median responses of approximately 1 and 1.5 standard deviations respectively, which are similar to the baseline peak of 1.4 standard deviation. We note that changing the Credit order has made the stress regime IRF be insignificant but the general hump-shape remains the same. 
Importantly, the qualitative result of a positive stress-regime response and a near-zero normal-regime response is invariant to the ordering.


\subsubsection*{Credit response}

Like the GDP results, the credit response to a regulatory capital shock displays also considerably a large difference between the stress and normal regimes across all specifications. We see that the stress regime response initially stays positive and then becomes negative from the second year onwards, albeit not significantly, before tapering off to 0. The cost of build-up leads to an opposite pattern, with credit response starting out at -0.6 standard deviations then turning positive in the second year onwards, now significantly, before tapering off the 0. 

Interestingly, the credit response in the normal regime starts out with a significant response, while it is insignificant initially in the stress regime . This is the opposite of what we see with GDPs response, which is significant in the stress regime. We caution against reading this as evidence of a regime difference in the credit channel itself, since the difference in significance status is not itself informative about the difference between the two responses. Nevertheless, it highlights that the credit response alone does not account for the magnitude of the output response, and this holds even in specifications where the credit response is sensitive to the ordering.
This points to the relevance of channels beyond credit supply. Specifically, banks may respond to a relaxation of capital requirements by adjusting the terms rather than the volume of lending. For instance banks might ease collateral requirements or extend maturities without necessarily increasing the total amount of credit they supply ceteris paribus. Since observed credit is an equilibrium quantity, an expansion in supply of this kind can be offset by contemporaneously weak demand and leave the aggregate flow largely unchanged \citep{becker2014cyclicality}. In addition, the capital release may operate through confidence and expectations effects: the macroprudential authority's decision to release the buffer signals that it stands ready to support the financial system, which may reduce precautionary behaviour.

Note that our results indicate a decoupling of credit and GDP in the normal regime. While the build-up of regulatory capital produces a modest contraction in credit (the normal-regime credit median is negative across specifications initially), the corresponding GDP response is essentially zero. This is consistent with the occasionally binding constraints interpretation that motivates the threshold specification. When borrowing constraints are slack (as they are by construction in the normal regime) the economy can substitute away from bank credit through alternative financing channels such as capital markets, retained earnings, or trade credit. A moderate tightening of bank capital requirements therefore does not materially reduce aggregate output. The decoupling breaks down in the stress regime, where constraints bind and the financial accelerator is active, which is where the release of regulatory capital matters.


Taken together, the impulse response results deliver two main conclusions. First, the GDP effect of regulatory capital shocks is strongly regime-dependent, with a positive and robust response to capital release in the stress regime and a negligible build-up cost in the normal regime. This finding is invariant to the Cholesky ordering, the sample period, and the definition of the credit variable, and it provides direct empirical support for the macroprudential rationale of the CCyB. Second, the credit response is more sensitive across specifications, which suggests that the aggregate credit flow variable captures only part of the transmission mechanism. The combination of these two findings implies that the benefits of CCyB release in periods of systemic stress extend beyond the narrow credit supply channel and operate through a broader set of macro-financial linkages.

\section{Conclusion}
This paper has used a Threshold Bayesian VAR to investigate whether the macroeconomic effects of capital regulation are regime-dependent, applying the framework to Hungarian monthly data. The FISS financial stress indicator governs the regime switch, linking the model to both the occasionally binding constraints literature \citep{brunnermeier_macroeconomic_2014,alessandri_financial_2017} and operational framework for CCyB release \citep{detken_operationalising_2014}. 

The results deliver strong evidence of the intended asymmetry in the effects of capital regulation. In the stress regime, releasing regulatory capital produces a positive and persistent GDP response, remaining positive for roughly twenty months. In the normal regime, the cost of building up capital is economically negligible. This combination is consistent with the design principles of the CCyB \citep{bank_of_england_financial_policy_committee_financial_2023,aikman_would_2019} and with micro-level evidence from Spain’s dynamic provisioning framework \citep{jimenez_macroprudential_2017}. The specification excluding unsecured consumer lending yields a cleaner credit signal, consistent with the hypothesis that CCyB release operates primarily through secured and corporate lending channels \citep{dursun-de_neef_countercyclical_2023,auer_countercyclical_2022}. 

There are seeral directions for future research. First, the recursive Cholesky identification could be complemented by sign restrictions or narrative approaches. Second, the two-regime specification may not capture the full complexity of occasionally binding constraints, and extensions to multiple thresholds or smooth transition models could highlight transmission channels especially for the build-up. Finally, complementing the aggregate analysis with bank-level panel evidence would provide a more complete picture of the transmission mechanism.

We note that where data availability permits, quantile VAR methods \citep{chavleishvili_forecasting_2024} remain the preferred framework for direct inference on conditional quantiles; the TBVAR is best understood as a practical compromise for settings where such methods are infeasible. Notwithstanding these caveats, the findings carry clear policy implications. The evidence that releasing capital requirements during financial stress supports GDP growth, without a corresponding cost during the accumulation phase, provides direct empirical justification for the countercyclical operation of the capital buffer. For small open economies with limited data histories, the TBVAR offers a tractable framework for assessing the regime-dependent effects of macroprudential instruments.

\pagebreak

\bibliographystyle{chicago}
\bibliography{reference.bib}

@article{szendrei_revisiting_2023,
    title = {Revisiting vulnerable growth in the euro area: {Identifying} the role of financial conditions in the distribution},
    volume = {223},
    shorttitle = {Revisiting vulnerable growth in the euro area},
    url = {https://www.sciencedirect.com/science/article/pii/S0165176523000150},
    urldate = {2026-02-25},
    journal = {Economics Letters},
    author = {Szendrei, Tibor and Varga, Katalin},
    year = {2023},
    pages = {110990},
}

@techreport{chavleishvili_risk_2021,
    title = {The risk management approach to macro-prudential policy},
    url = {https://www.econstor.eu/handle/10419/237704},
    urldate = {2026-02-25},
    institution = {ECB working paper},
    author = {Chavleishvili, Sulkhan and Engle, Robert F. and Fahr, Stephan and Kremer, Manfred and Manganelli, Simone and Schwaab, Bernd},
    year = {2021},
}

@article{galan_benefits_2024,
    title = {The benefits are at the tail: uncovering the impact of macroprudential policy on growth-at-risk},
    volume = {74},
    shorttitle = {The benefits are at the tail},
    url = {https://www.sciencedirect.com/science/article/pii/S1572308920301340},
    urldate = {2026-02-25},
    journal = {Journal of Financial Stability},
    publisher = {Elsevier},
    author = {Galán, Jorge E.},
    year = {2024},
    pages = {100831},
}

@article{skrinjaric_impact_2025,
    title = {Impact of {Macroprudential} {Policy} on {Economic} {Growth}: {A} {Survey} on {Growth}‐at‐{Risk} {Approach}},
    issn = {0950-0804, 1467-6419},
    shorttitle = {Impact of {Macroprudential} {Policy} on {Economic} {Growth}},
    url = {https://onlinelibrary.wiley.com/doi/10.1111/joes.70052},
    doi = {10.1111/joes.70052},
    language = {en},
    urldate = {2026-02-25},
    journal = {Journal of Economic Surveys},
    author = {Škrinjarić, Tihana},
    month = dec,
    year = {2025},
    pages = {joes.70052},
}

@article{chavleishvili_forecasting_2024,
    title = {Forecasting and stress testing with quantile vector autoregression},
    volume = {39},
    issn = {0883-7252, 1099-1255},
    url = {https://onlinelibrary.wiley.com/doi/10.1002/jae.3009},
    doi = {10.1002/jae.3009},
    language = {en},
    number = {1},
    urldate = {2026-02-25},
    journal = {Journal of Applied Econometrics},
    author = {Chavleishvili, Sulkhan and Manganelli, Simone},
    month = jan,
    year = {2024},
    pages = {66--85},
}

@article{caldara_understanding_2020,
    title = {Understanding growth-at-risk: {A} {Markov}-switching approach},
    shorttitle = {Understanding growth-at-risk},
    url = {https://real-time-economics.org/wp-content/uploads/2021/04/borda.pdf},
    urldate = {2026-02-25},
    journal = {Available at SSRN},
    author = {Caldara, Dario and Cascaldi-Garcia, Danilo and Cuba-Borda, Pablo and Loria, Francesca},
    year = {2020},
}

@article{carriero_capturing_2024,
    title = {Capturing {Macro}‐{Economic} {Tail} {Risks} with {Bayesian} {Vector} {Autoregressions}},
    volume = {56},
    issn = {0022-2879, 1538-4616},
    url = {https://onlinelibrary.wiley.com/doi/10.1111/jmcb.13121},
    doi = {10.1111/jmcb.13121},
    language = {en},
    number = {5},
    urldate = {2026-02-25},
    journal = {Journal of Money, Credit and Banking},
    author = {Carriero, Andrea and Clark, Todd E. and Marcellino, Massimiliano},
    month = aug,
    year = {2024},
    pages = {1099--1127},
}

@article{varga_non-stationary_2025,
    title = {Non-stationary financial risk factors and macroeconomic vulnerability for the {UK}},
    volume = {97},
    url = {https://www.sciencedirect.com/science/article/pii/S1057521924007981},
    urldate = {2026-02-25},
    journal = {International Review of Financial Analysis},
    publisher = {Elsevier},
    author = {Varga, Katalin and Szendrei, Tibor},
    year = {2025},
    pages = {103866},
}

@article{jimenez_macroprudential_2017,
    title = {Macroprudential {Policy}, {Countercyclical} {Bank} {Capital} {Buffers}, and {Credit} {Supply}: {Evidence} from the {Spanish} {Dynamic} {Provisioning} {Experiments}},
    volume = {125},
    issn = {0022-3808, 1537-534X},
    shorttitle = {Macroprudential {Policy}, {Countercyclical} {Bank} {Capital} {Buffers}, and {Credit} {Supply}},
    url = {https://www.journals.uchicago.edu/doi/10.1086/694289},
    doi = {10.1086/694289},
    language = {en},
    number = {6},
    urldate = {2026-02-25},
    journal = {Journal of Political Economy},
    author = {Jiménez, Gabriel and Ongena, Steven and Peydró, José-Luis and Saurina, Jesús},
    month = dec,
    year = {2017},
    pages = {2126--2177},
}

@techreport{bank_of_england_financial_policy_committee_financial_2023,
    type = {Policy {Statement}},
    title = {The {Financial} {Policy} {Committee}'s {Approach} to {Setting} the {Countercyclical} {Capital} {Buffer}},
    url = {https://www.bankofengland.co.uk/paper/2023/ps/the-financial-policy-committees-approach-to-setting-the-countercyclical-capital-buffer},
    institution = {Bank of England},
    author = {{Bank of England, Financial Policy Committee}},
    month = jul,
    year = {2023},
}

@article{aikman_would_2019,
    title = {Would macroprudential regulation have prevented the last crisis?},
    volume = {33},
    url = {https://www.aeaweb.org/articles?id=10.1257/jep.33.1.107},
    number = {1},
    urldate = {2026-02-25},
    journal = {Journal of Economic Perspectives},
    publisher = {American Economic Association 2014 Broadway, Suite 305, Nashville, TN 37203-2418},
    author = {Aikman, David and Bridges, Jonathan and Kashyap, Anil and Siegert, Caspar},
    year = {2019},
    pages = {107--130},
}

@article{carlier_vector_2016,
    title = {Vector quantile regression: {An} optimal transport approach},
    volume = {44},
    issn = {0090-5364, 2168-8966},
    shorttitle = {Vector quantile regression},
    url = {https://projecteuclid.org/journals/annals-of-statistics/volume-44/issue-3/Vector-quantile-regression-An-optimal-transport-approach/10.1214/15-AOS1401.full},
    doi = {10.1214/15-AOS1401},
    language = {en},
    number = {3},
    urldate = {2026-02-26},
    journal = {The Annals of Statistics},
    publisher = {Institute of Mathematical Statistics},
    author = {Carlier, Guillaume and Chernozhukov, Victor and Galichon, Alfred},
    month = jun,
    year = {2016},
    pages = {1165--1192},
}

@misc{szendrei_roof_2026,
    title = {A {Roof} {Over} {Risk}: {A} {House} {Price}-at-{Risk} {Framework} for {Hungary}},
    shorttitle = {A {Roof} {Over} {Risk}},
    url = {http://arxiv.org/abs/2602.18592},
    doi = {10.48550/arXiv.2602.18592},
    urldate = {2026-02-26},
    publisher = {arXiv},
    author = {Szendrei, Tibor and Vágó, Nikolett and Varga, Katalin},
    month = feb,
    year = {2026},
    note = {arXiv:2602.18592 [econ]},
}

@article{dursun-de_neef_countercyclical_2023,
    title = {Countercyclical capital buffers and credit supply: evidence from the {COVID}-19 crisis},
    volume = {154},
    shorttitle = {Countercyclical capital buffers and credit supply},
    url = {https://www.sciencedirect.com/science/article/pii/S0378426623001358},
    urldate = {2026-02-26},
    journal = {Journal of Banking \& Finance},
    publisher = {Elsevier},
    author = {Dursun-de Neef, H. {\"O}zlem and Schandlbauer, Alexander and Wittig, Colin},
    year = {2023},
    pages = {106930},
}

@article{auer_countercyclical_2022,
    title = {The countercyclical capital buffer and the composition of bank lending},
    volume = {52},
    url = {https://www.sciencedirect.com/science/article/pii/S1042957322000183},
    urldate = {2026-02-26},
    journal = {Journal of Financial Intermediation},
    publisher = {Elsevier},
    author = {Auer, Raphael and Matyunina, Alexandra and Ongena, Steven},
    year = {2022},
    pages = {100965},
}

@article{adrian_vulnerable_2019,
    title = {Vulnerable growth},
    volume = {109},
    url = {https://www.aeaweb.org/articles?id=10.1257/aer.20161923},
    number = {4},
    urldate = {2026-03-03},
    journal = {American Economic Review},
    publisher = {American Economic Association 2014 Broadway, Suite 305, Nashville, TN 37203},
    author = {Adrian, Tobias and Boyarchenko, Nina and Giannone, Domenico},
    year = {2019},
    pages = {1263--1289},
}

@article{figueres_vulnerable_2020,
    title = {Vulnerable growth in the euro area: {Measuring} the financial conditions},
    volume = {191},
    shorttitle = {Vulnerable growth in the euro area},
    url = {https://www.sciencedirect.com/science/article/pii/S016517652030104X},
    urldate = {2026-03-03},
    journal = {Economics Letters},
    publisher = {Elsevier},
    author = {Figueres, Juan Manuel and Jarociński, Marek},
    year = {2020},
    pages = {109126},
}

@article{koop_nowcasting_2021,
    title = {Nowcasting ‘true’monthly {US} {GDP} during the pandemic},
    volume = {256},
    url = {https://www.cambridge.org/core/journals/national-institute-economic-review/article/nowcasting-true-monthly-us-gdp-during-the-pandemic/FC9DF95AC12FEF6759E87AD6B97C6B4D},
    urldate = {2026-03-03},
    journal = {National Institute Economic Review},
    publisher = {Cambridge University Press},
    author = {Koop, Gary and McIntyre, Stuart and Mitchell, James and Poon, Aubrey},
    year = {2021},
    pages = {44--70},
}

@article{koop_reconciled_2023,
    title = {Reconciled {Estimates} of {Monthly} {GDP} in the {United} {States}},
    volume = {41},
    issn = {0735-0015, 1537-2707},
    url = {https://www.tandfonline.com/doi/full/10.1080/07350015.2022.2044336},
    doi = {10.1080/07350015.2022.2044336},
    language = {en},
    number = {2},
    urldate = {2026-03-03},
    journal = {Journal of Business \& Economic Statistics},
    author = {Koop, Gary and McIntyre, Stuart and Mitchell, James and Poon, Aubrey},
    month = apr,
    year = {2023},
    pages = {563--577},
}

@article{huber_nowcasting_2023,
    title = {Nowcasting in a pandemic using non-parametric mixed frequency {VARs}},
    volume = {232},
    url = {https://www.sciencedirect.com/science/article/pii/S0304407620303936},
    number = {1},
    urldate = {2026-03-03},
    journal = {Journal of Econometrics},
    publisher = {Elsevier},
    author = {Huber, Florian and Koop, Gary and Onorante, Luca and Pfarrhofer, Michael and Schreiner, Josef},
    year = {2023},
    pages = {52--69},
}

@techreport{schorfheide_real-time_2021,
    title = {Real-time forecasting with a (standard) mixed-frequency {VAR} during a pandemic},
    url = {https://www.nber.org/papers/w29535},
    urldate = {2026-03-03},
    institution = {National Bureau of Economic Research},
    author = {Schorfheide, Frank and Song, Dongho},
    year = {2021},
}

@article{szendrei_fiss_2020,
    title = {{FISS} - {A} {Factor}-based {Index} of {Systemic} {Stress} in the {Financial} {System}},
    volume = {79},
    url = {https://EconPapers.repec.org/RePEc:bkr:journl:v:79:y:2020:i:1:p:3-34},
    doi = {10.31477/rjmf.202001.03},
    number = {1},
    urldate = {2026-03-03},
    journal = {Russian Journal of Money and Finance},
    publisher = {Bank of Russia},
    author = {Szendrei, Tibor and Varga, Katalin},
    year = {2020},
    pages = {3--34},
}

@article{brunnermeier_macroeconomic_2014,
    title = {A macroeconomic model with a financial sector},
    volume = {104},
    url = {https://www.aeaweb.org/articles?id=10.1257/aer.104.2.379},
    number = {2},
    urldate = {2026-03-03},
    journal = {American Economic Review},
    publisher = {American Economic Association 2014 Broadway, Suite 305, Nashville, TN 37203},
    author = {Brunnermeier, Markus K. and Sannikov, Yuliy},
    year = {2014},
    pages = {379--421},
}

@article{he_model_2012,
    title = {A model of capital and crises},
    volume = {79},
    url = {https://academic.oup.com/restud/article-abstract/79/2/735/1533631},
    number = {2},
    urldate = {2026-03-03},
    journal = {The Review of Economic Studies},
    publisher = {Oxford University Press},
    author = {He, Zhigu and Krishnamurthy, Arvind},
    year = {2012},
    pages = {735--777},
}

@article{aruoba_piecewise-linear_2021,
    title = {Piecewise-linear approximations and filtering for {DSGE} models with occasionally-binding constraints},
    volume = {41},
    url = {https://www.sciencedirect.com/science/article/pii/S1094202520301149},
    urldate = {2026-03-03},
    journal = {Review of Economic Dynamics},
    publisher = {Elsevier},
    author = {Aruoba, S. Borağan and Cuba-Borda, Pablo and Higa-Flores, Kenji and Schorfheide, Frank and Villalvazo, Sergio},
    year = {2021},
    pages = {96--120},
}

@article{aruoba_svars_2022,
    title = {{SVARs} with occasionally-binding constraints},
    volume = {231},
    url = {https://www.sciencedirect.com/science/article/pii/S0304407621002487},
    number = {2},
    urldate = {2026-03-03},
    journal = {Journal of Econometrics},
    publisher = {Elsevier},
    author = {Aruoba, S. Borağan and Mlikota, Marko and Schorfheide, Frank and Villalvazo, Sergio},
    year = {2022},
    pages = {477--499},
}

@article{alessandri_financial_2017,
    title = {Financial conditions and density forecasts for {US} output and inflation},
    volume = {24},
    url = {https://www.sciencedirect.com/science/article/pii/S1094202517300042},
    urldate = {2026-03-03},
    journal = {Review of Economic Dynamics},
    publisher = {Elsevier},
    author = {Alessandri, Piergiorgio and Mumtaz, Haroon},
    year = {2017},
    pages = {66--78},
}

@article{bernanke_measuring_2005,
    title = {Measuring the effects of monetary policy: a factor-augmented vector autoregressive ({FAVAR}) approach},
    volume = {120},
    shorttitle = {Measuring the effects of monetary policy},
    url = {https://academic.oup.com/qje/article-abstract/120/1/387/1931468},
    number = {1},
    urldate = {2026-03-03},
    journal = {The Quarterly journal of economics},
    publisher = {MIT Press},
    author = {Bernanke, Ben S. and Boivin, Jean and Eliasz, Piotr},
    year = {2005},
    pages = {387--422},
}

@book{european_systemic_risk_board_improvements_2024,
    address = {LU},
    title = {Improvements to the {ESRB} macroprudential stance framework: report by the {Contact} {Group} on {Macroprudential} {Stance} of the {ESRB}’s {Instruments} {Working} {Group} ({IWG})},
    shorttitle = {Improvements to the {ESRB} macroprudential stance framework},
    url = {https://data.europa.eu/doi/10.2849/21350},
    doi = {10.2849/21350},
    language = {eng},
    urldate = {2026-03-02},
    publisher = {Publications Office},
    author = {{European Systemic Risk Board.}},
    year = {2024},
}

@article{chen_bayesian_1995,
    title = {{BAYESIAN} {INFERENCE} {OF} {THRESHOLD} {AUTOREGRESSIVE} {MODELS}},
    volume = {16},
    copyright = {http://onlinelibrary.wiley.com/termsAndConditions\#vor},
    issn = {0143-9782, 1467-9892},
    url = {https://onlinelibrary.wiley.com/doi/10.1111/j.1467-9892.1995.tb00248.x},
    doi = {10.1111/j.1467-9892.1995.tb00248.x},
    language = {en},
    number = {5},
    urldate = {2026-03-04},
    journal = {Journal of Time Series Analysis},
    author = {Chen, Cathy W. S. and Lee, Jack C.},
    month = sep,
    year = {1995},
    pages = {483--492},
}

@article{koop_impulse_1996,
    title = {Impulse response analysis in nonlinear multivariate models},
    volume = {74},
    url = {https://www.sciencedirect.com/science/article/pii/0304407695017534},
    number = {1},
    urldate = {2026-03-04},
    journal = {Journal of econometrics},
    publisher = {Elsevier},
    author = {Koop, Gary and Pesaran, M. Hashem and Potter, Simon M.},
    year = {1996},
    pages = {119--147},
}

@article{hubrich_financial_2015,
    title = {Financial stress and economic dynamics: {The} transmission of crises},
    volume = {70},
    shorttitle = {Financial stress and economic dynamics},
    url = {https://www.sciencedirect.com/science/article/pii/S030439321400155X},
    urldate = {2026-03-04},
    journal = {Journal of Monetary Economics},
    publisher = {Elsevier},
    author = {Hubrich, Kirstin and Tetlow, Robert J.},
    year = {2015},
    pages = {100--115},
}

@article{mumtaz_transmission_2015,
    title = {{THE} {TRANSMISSION} {MECHANISM} {IN} {GOOD} {AND} {BAD} {TIMES}},
    volume = {56},
    copyright = {http://onlinelibrary.wiley.com/termsAndConditions\#vor},
    issn = {0020-6598, 1468-2354},
    url = {https://onlinelibrary.wiley.com/doi/10.1111/iere.12136},
    doi = {10.1111/iere.12136},
    language = {en},
    number = {4},
    urldate = {2026-03-04},
    journal = {International Economic Review},
    author = {Mumtaz, Haroon and Surico, Paolo},
    month = nov,
    year = {2015},
    pages = {1237--1260},
}

@article{potter_nonlinear_2000,
    title = {Nonlinear impulse response functions},
    volume = {24},
    url = {https://www.sciencedirect.com/science/article/pii/S0165188999000135},
    number = {10},
    urldate = {2026-03-04},
    journal = {Journal of Economic Dynamics and Control},
    publisher = {Elsevier},
    author = {Potter, Simon M.},
    year = {2000},
    pages = {1425--1446},
}

@article{fernandes_smoothing_2021,
    title = {Smoothing {Quantile} {Regressions}},
    volume = {39},
    issn = {0735-0015, 1537-2707},
    url = {https://www.tandfonline.com/doi/full/10.1080/07350015.2019.1660177},
    doi = {10.1080/07350015.2019.1660177},
    language = {en},
    number = {1},
    urldate = {2026-03-05},
    journal = {Journal of Business \& Economic Statistics},
    author = {Fernandes, Marcelo and Guerre, Emmanuel and Horta, Eduardo},
    month = jan,
    year = {2021},
    pages = {338--357},
}

@article{sims_macroeconomics_1980,
    title = {Macroeconomics and reality},
    url = {https://www.jstor.org/stable/1912017},
    urldate = {2026-03-05},
    journal = {Econometrica: journal of the Econometric Society},
    publisher = {JSTOR},
    author = {Sims, Christopher A.},
    year = {1980},
    pages = {1--48},
}

@article{alessandri_financial_2019,
    title = {Financial regimes and uncertainty shocks},
    volume = {101},
    url = {https://www.sciencedirect.com/science/article/pii/S0304393218302745},
    urldate = {2026-03-05},
    journal = {Journal of Monetary Economics},
    publisher = {Elsevier},
    author = {Alessandri, Piergiorgio and Mumtaz, Haroon},
    year = {2019},
    pages = {31--46},
}

@article{detken_operationalising_2014,
    title = {Operationalising the countercyclical capital buffer: indicator selection, threshold identification and calibration options},
    shorttitle = {Operationalising the countercyclical capital buffer},
    url = {https://papers.ssrn.com/sol3/papers.cfm?abstract_id=3723336},
    number = {2014/5},
    urldate = {2026-03-05},
    journal = {ESRB: Occasional Paper Series},
    author = {Detken, Carsten and Weeken, Olaf and Alessi, Lucia and Bonfim, Diana and Boucinha, Miguel and Castro, Christian and Frontczak, Sebastian and Giordana, Gaston and Giese, Julia and Wildmann, Nadya},
    year = {2014},
}

@article{iacoviello_house_2005,
    title = {House prices, borrowing constraints, and monetary policy in the business cycle},
    volume = {95},
    url = {https://www.aeaweb.org/articles?id=10.1257/0002828054201477},
    number = {3},
    urldate = {2026-03-09},
    journal = {American economic review},
    publisher = {American Economic Association},
    author = {Iacoviello, Matteo},
    year = {2005},
    pages = {739--764},
}

@techreport{magyar_nemzeti_bank_financial_2022,
    address = {Budapest},
    title = {Financial {Stability} {Report}: {November} 2022},
    url = {https://www.mnb.hu/letoltes/financial-stability-report-november-2022.pdf},
    institution = {Magyar Nemzeti Bank},
    author = {{Magyar Nemzeti Bank}},
    month = nov,
    year = {2022},
}

@techreport{imf2024hungary_articleiv,
  author       = {{International Monetary Fund}},
  title        = {Hungary: 2024 Article IV Consultation},
  institution  = {International Monetary Fund},
  series       = {IMF Staff Country Reports},
  number       = {2024/268},
  year         = {2024},
  month        = aug,
  address      = {Washington, DC},
  doi          = {10.5089/9798400283703.002},
  url          = {https://www.imf.org/en/-/media/files/publications/cr/2024/english/1hunea2024001-print-pdf.pdf},
  note         = {Accessed: 2026-03-09}
}

@article{becker2014cyclicality,
  title={Cyclicality of credit supply: Firm level evidence},
  author={Becker, Bo and Ivashina, Victoria},
  journal={Journal of Monetary Economics},
  volume={62},
  pages={76--93},
  year={2014},
  publisher={Elsevier}
}

@article{aikman2019credit,
  title={Credit, capital and crises: a GDP-at-Risk approach},
  author={Aikman, David and Bridges, Jonathan and Hacioglu Hoke, Sinem and O'Neill, Cian and Raja, Akash},
  year={2019},
  publisher={Bank of England Working Paper}
}

@article{franta2020effects,
  title={On the effects of macroprudential policies on Growth-at-Risk},
  author={Franta, Michal and Gambacorta, Leonardo},
  journal={Economics Letters},
  volume={196},
  pages={109501},
  year={2020},
  publisher={Elsevier}
}

\pagebreak


\appendix 
\section{Appendix}
\subsection{Monthly GDP}
To obtain the monthly GDP values, we start from quarterly GDP values based on expenditure and income approaches. The key is to consider these series as noisy observations of true GDP. Given that these measures are noisy approximations, \citet{koop_reconciled_2023} employs a noise restriction based on the assumption that variance of true GDP is less than the variance of its noisy observations.\footnote{The underlying dataset and details of the priors can be found in the Technical Appendix.} Our quarterly frequency equations are the following:

\begin{equation} \label{eq:monthylGDP1}
\begin{split}
    \begin{bmatrix}
        GDP_{P,t}\\GDP_{E,t}
    \end{bmatrix}&=1_{2 \times 1} \mathbf{GDP}_t+
    \begin{bmatrix}
        \varepsilon_{P,t} \\ \varepsilon_{E,t}
    \end{bmatrix} \\
    GDP_t&=\rho \mathbf{GDP}_{t-1}+\varepsilon_{G,t}
\end{split}
\end{equation}

\noindent where $GDP_{P,t}$ is production-based GDP, $GDP_{E,t}$ is expenditure-based GDP, and $\mathbf{GDP}_{t}$ is true GDP. Since $\mathbf{GDP}_{t}$ is unobserved, the model in equation (\ref{eq:monthylGDP1}) is not identified without further restrictions since there is no unique way to separate the true signal from the measurement noise in $GDP_{P,t}$ and $GDP_{E,t}$. To see why, consider what the two equations in (\ref{eq:monthylGDP1}) tell us: both income- and expenditure-based GDP are assumed to be noisy readings of the same underlying truth. If the noise is large relative to the true variation in GDP, we would expect the observed series to be more volatile than the true series. Conversely, if there is genuine information about the economy (i.e. news rather than noise) then the observed series may actually be less volatile than true GDP. This intuition motivates using the relative volatility of true GDP through the variance ratio:

\begin{equation}
    \xi_i=\frac{var(\mathbf{GDP})}{var(GDP_i)}
\end{equation}

The above reparametrisation reflects that the error in measurement hypothesis, namely that income- and expenditure-based GDP is equal to true GDP plus measurement error. When $\xi_i<1$, true GDP is less volatile than its observed counterpart, consistent with the noise hypothesis. When $\xi_i>1$, the reverse is true. Rather than fixing $\xi_i$ to a single value (which would be a strong assumption) we follow \citet{koop_reconciled_2023} in treating it as a quantity to be estimated, with priors defined as intervals that reflect our uncertainty about whether noise or news dominates.

The mixed frequency model can be summarised as:

\begin{equation}
\begin{split}
    Y_t&=(X'_t,U_t,GDP_t,GDP_{P,t},GDP_{E,t}) \\
    Y_t^Q&=\Delta_3\ln (Y_t) 
\end{split}
\end{equation}

\noindent where $Y_t^Q$ is the quarterly variable observed every third month, $U_t$ is the unemployment rate, that depends on $GDP$ but not on $GDP_P$ or $GDP_E$, and $X_t'$ is a set of monthly explanatory variables. The data used to construct the Monthly GDP are the following:

\begin{itemize}
    \item Monthly variables
        \begin{enumerate}
            \item Industrial Production Index, monthly, volume or constant price value index, SA, source: HCSO, Hungarian Central Statistical Office
            \item Retail Sales, monthly, total, volume, calendar adjusted, index, corresponding period of the previous year = 100, source: HCSO
            \item CPI, core, price index, SA, source: HCSO
            \item BUX Index, source: Refinitiv
            \item Unemployment rate, monthly, percentage, SA, source: Refinitive/ OECD
            \item 3-Month Benchmark Bond, monthly, interest rate, source: Refinitiv
            \item 10-year Benchmark Bond, monthly, interest rate, source: Refinitiv
            \item Construction Output, total, volume, index, corresponding period of the previous year = 100, source: HCSO
        \end{enumerate}
    \item Quarterly variables
    \begin{enumerate}
        \item GDP real, production based (HCSO)
        \item GDP real, expenditure based (HCSO)
    \end{enumerate}
    
\end{itemize}


\subsection{Prior for the quarterly VAR containing only GDP variables}

The prior distribution is restricted, so the implied variance ratios satisfy $0.36 < \xi_E, \xi_P < 1.15$. The priors are specified as follows:

\begin{enumerate}
    \item $a_{21}, a_{31}, a_{32} \sim N(0,10)$,
    \item $\mu \sim N(0,100)$ and $b_{11} \sim N(0,10)$,
    \item $\sigma^2_{GG}, \sigma^2_{EE}, \sigma^2_{PP} \sim IG(3.8, 8.4)$, where the inverse gamma prior has mean 3 and variance 5.
\end{enumerate}

\subsection{Prior for the quarterly VAR containing unemployment and GDP variables}

As in the previous specification, the prior is bounded to ensure that the implied variance ratios satisfy  $0.36 < \xi_E, \xi_P < 1.15$. The prior distributions are given by:

\begin{enumerate}
    \item $Taa_{21} \sim N(0.5,1)$, $a_{32}, a_{42} \sim N(-1,0.1)$ and $a_{43} \sim N(0,1)$,
    \item $\mu, \mu_b \sim N(0,100)$, while $b_{11}, b_{12}, b_{21}, b_{22} \sim N(0,10)$,
    \item $\sigma^2_{UU}, \sigma^2_{GG}, \sigma^2_{EE}, \sigma^2_{PP} \sim IG(3.8, 8.4)$, with an inverse gamma prior whose mean equals 3 and variance equals 5.
\end{enumerate}

The structural form of the MF-VAR system can be written as:

\[
\begin{bmatrix}
1       & 0       &   0     & \cdots       & 0      & 0  & 0        & 0 \\
a_{2,1}  & 1       &   0     & \cdots       & 0      & 0  & 0        & 0 \\
a_{3,1}  & a_{3,2}  & 1       & \cdots       & 0      & 0  & 0        & 0 \\
a_{4,1}  & a_{4,2}  & a_{4,3}  & \cdots       & 0      & 0  & 0        & 0 \\
\vdots  & \vdots  & \vdots  & \ddots       & \vdots & \vdots&\vdots&\vdots\\
a_{8,1}  & a_{8,2}  & a_{8,3}  & \cdots       & 1      & 0  & 0        & 0 \\
a_{9,1}  & a_{9,2}  & a_{9,3}  & \cdots       & a_{9,8} & 1  & 0        & 0 \\
a_{10,1} & a_{10,2} & a_{10,3} & \cdots       & 0      & -1 & 1        & 0 \\
a_{11,1} & a_{11,2} & a_{11,3} & \cdots       & 0      & -1 &-a_{11,10}  & 1
\end{bmatrix}
\begin{bmatrix}
IndPro_t \\
RetSal_t \\
Buxind_t \\
CPI_t \\
Const_t \\
Bond10Y_t \\
Bond3M_t \\
Unemp_t \\
GDP_t \\
GDP_{E,t} \\
GDP_{P,t}
\end{bmatrix}
\]

We use the following notation: $\hat{a} = (a_{21}, a_{31}, \cdots a_{116}, a_{117})$ denote the vector collecting the corresponding coefficients in matrix $A$. Let $\Tilde{a}$ represent the remaining coefficients in $A$ together with all free coefficients in $B$ and the intercept terms in the MF-VAR.

The term $\sigma_{ii}^2$ denotes the error variance in equation $i$.

The prior distributions are specified as:

\begin{enumerate}
    \item $a_{1110} \sim N(0,1)$,
    \item $\tilde{a} \sim DL(\alpha)$, where $\alpha$ is the hyperparameter governing the Dirichlet–Laplace prior and is set to $\alpha = 0.5$,
    \item $\tilde{a} \sim DL(\bar{\alpha})$, where $\bar{\alpha}$ denotes the
    corresponding hyperparameter and is fixed at $\bar{\alpha} = 0.5$
\end{enumerate}

The prior is bounded, so that $0.36 < \xi_E, \xi_I < 1.15.
$ 

\end{document}